\documentclass[fleqn,usenatbib]{mnras}

\usepackage{newtxtext,newtxmath}

\usepackage[T1]{fontenc}
\usepackage{ae,aecompl}

\usepackage{graphicx}	
\usepackage{amsmath}	
\usepackage{amssymb}	

\title[Effects of diffusion on hot subdwarf formation]{The effects of diffusion in hot subdwarf progenitors from the common envelope channel}

\author[Byrne et. al]{
Conor M. Byrne,$^{1,2}$\thanks{E-mail: cmb@arm.ac.uk (CMB)}
C. Simon Jeffery,$^{1,2}$
Christopher A. Tout$^{3}$
and Haili Hu$^{4}$
\\
$^{1}$Armagh Observatory \& Planetarium, College Hill, Armagh BT61 9DG, UK\\
$^{2}$School of Physics, Trinity College Dublin, College Green, Dublin 2, Ireland\\
$^{3}$Institute of Astronomy, University of Cambridge, Madingley Road, Cambridge CB3 0HA, UK\\
$^{4}$Earth Science Group, SRON Netherlands Institute for Space Research, Sorbonnelaan 2, 3584 CA Utrecht, The Netherlands
}

\date{Accepted 2018 January 16. Received 2018 January 16; in original form 2017 August 7}

\pubyear{2017}

\begin{document}
\label{firstpage}
\pagerange{\pageref{firstpage}--\pageref{lastpage}}
\maketitle

\begin{abstract}
Diffusion of elements in the atmosphere and envelope of a star can drastically alter its surface composition,  leading to extreme chemical peculiarities. We consider the case of hot subdwarfs, where surface helium abundances range from practically zero to almost 100 percent. Since hot subdwarfs can form via a number of different evolution channels, a key question concerns how the formation mechanism is connected to the present surface chemistry. A sequence of extreme horizontal branch star models was generated by producing post-common envelope stars from red giants. Evolution was computed with {\tt MESA} from envelope ejection up to core-helium ignition. Surface abundances were calculated at the zero-age horizontal branch for models  with and without diffusion. A number of simulations also included radiative levitation.  The goal was to study surface chemistry during evolution from cool giant to hot subdwarf and determine when the characteristic subdwarf surface is established. Only stars leaving the giant branch close to core-helium ignition become hydrogen-rich subdwarfs at the zero-age horizontal branch. Diffusion, including radiative levitation, depletes the initial surface helium in all cases. All subdwarf models rapidly become  more depleted  than observations allow. Surface abundances of other elements follow observed trends in general, but not in detail. Additional physics is required. 
\end{abstract}




\section{Introduction}

The abundances of chemical elements on the surface of a star depend on many different physical processes. These include mass transfer in a binary star system, accretion from a circumstellar disc, mixing of nucleosynthesised material from the interior to the surface, magnetic fields and atomic diffusion. If a star shows large deviations in abundances of certain elements from a standard or average star of its type, it may be classed as a chemically peculiar star. 

Atomic diffusion is a term used to describe a group of particle transport processes which act to modify the chemical structure of a star, provided the material is hydrodynamically stable. These processes are thermal diffusion, concentration diffusion, gravitational settling and radiative levitation. 

In most stellar atmospheres, one element (either H or He) typically dominates and the effects of concentration diffusion can be neglected. If a star has a steep pressure gradient, this dominates over thermal diffusion, so reducing the diffusion problem to a balance between the inward force of gravity and the outward force of radiation. The different forces acting on different elements lead to changes in the composition of the atmosphere and has been shown to explain different types of chemically peculiar stars, such as the Ap stars \citep{Michaud70}.

One particular group of stars which show many anomalous surface abundances are the hot subdwarf stars. These low-mass (about $ 0.5\,\mathrm{M}_\odot$), helium core burning stars can vary massively in appearance from atmospheres almost entirely comprised of hydrogen to those which are extremely helium-rich. Some subdwarfs show extremely anomalous abundances of lead, zirconium and other heavy elements \citep{Naslim11,Jeffery17}. This diverse population of stars is thus an ideal environment in which to test the treatment of diffusion in stellar evolution simulations. Diffusion is necessary to explain some of the unusual properties of these stars, such as the presence of pulsations, caused by Z-bump opacity due to radiative levitation of iron \citep{Charpinet97,Fontaine03} and nickel \citep{JefferySaio06b}.

Three main formation channels for hot subdwarfs have been identified \citep{Han02,Han03}. Hot subdwarfs may be formed by the merging of two low-mass white dwarfs, by stable mass transfer via Roche lobe overflow (RLOF) from a red giant to a low-mass binary companion or by unstable mass transfer between a red giant and a low-mass star which leads to the formation of a common envelope that is subsequently ejected. These three channels produce single hot subdwarfs, subdwarfs in long-period binary orbits and subdwarfs in short-period binary orbits respectively.

Computational modelling of double white dwarf mergers has shown that this can explain the formation of helium-rich hot subdwarfs \citep{Zhang12}. However this evolution channel would produce single hot subdwarfs and not binaries. At least one helium-rich subdwarf is known to be in a spectroscopic binary with an orbital period of $2.3$\,d, which indicates a post-common-envelope rather than a merger origin \citep{Naslim12}. The common envelope ejection is a poorly understood phase of evolution, and the exact outcome of such events is unclear. A recent review of common envelope evolution is given by \cite{IvanovaRev13}, while hot subdwarfs are discussed at length  by \citet{Heber16}.

What makes hot subdwarfs such an interesting area of research is the diversity of surface chemistries which exist in the same region of the HR diagram. Spectroscopic determinations of the surface helium abundances of these stars as a function of effective temperature reveal 2 distinct sequences of stars  as seen in fig. 6 of \citet{Nemeth12}. The majority follow a trend of increasing helium abundance with increasing temperature. However, a small number of hot subdwarfs fit into a more helium-deficient sequence. The observations also illustrate that a majority of subdwarf B type stars (sdBs) are helium-deficient, with a small number being helium-rich. 
The concentration of many different types of stars into a small region of the HR diagram indicates that these populations may have had different evolutionary histories. 
 
Studies of the evolution of post-common-envelope hot subdwarfs have been carried out by, for example, \citet{Xiong17}. However the effects of atomic diffusion have not always been considered. Clear understanding of the results of diffusion processes in these stars is needed in order to find the conditions in which the surfaces of hot subdwarfs become hydrogen-rich or helium-rich.

We carry out an investigation of both of these key elements of hot subdwarf physics (namely diffusion and evolution) in an attempt to quantify the effects of the history of these objects on their present state and attempt to link this to the different populations of observed hot subdwarf stars.

\section{Methods}
\label{sec:methods} 
Version 7624 of the \texttt{MESA} stellar evolution code (Modules for Experiments in Stellar Astrophysics; \cite{Paxton11,Paxton13,Paxton15}) was used to carry out this research. {\texttt{MESA}} was chosen because it is a robust code capable of approximating the evolution through helium flashes, a key step in the evolution of hot subdwarfs which populate the extreme horizontal branch (EHB) and in which helium core burning has started. 

Input physics parameters were chosen to be similar to that of other recent work with {\texttt{MESA}} on hot subdwarfs \citep{Xiong17, Schindler15}, that is starting with a Population I atomic composition ($Z=0.02$ and $X=0.7$) with other important parameters listed in Table~\ref{tab:parameters}, including recent calibrations of the mixing length parameter in {\texttt{MESA}} \citep{Stancliffe16}. Because the phase of evolution examined during this work was the transition from the red giant branch (RGB) to the extreme horizontal branch (EHB) at the onset of core helium burning, type I opacity tables were used. These simulations were carried out under the assumption that the star is sufficiently stable in this transition phase to allow atomic diffusion to operate.

\begin{table}
	\centering
	\caption{Key physics choices made for the models produced}
	\label{tab:parameters}
	\begin{tabular}{lr}
		\hline
		Parameter & Value\\
		\hline \hline
		Opacity & OP, Type I \citep{OP1,OP2}\\ 
		$\alpha_{\mathrm{MLT}}$ & 1.9\\ 
		Composition  & \cite{Grevesse98}\\
		Mass loss & \texttt{relax\_mass} (during CE ejection only)\\ 
		Convection & Schwarzchild criterion\\ \hline
	 	 & None (basic models)\\ 
		 & Thermal Diffusion (standard \& complete)\\
		Diffusion & Concentration Diffusion (standard \& complete)\\
		 & Gravitational Settling (standard \& complete)\\ 
		 & Radiative Levitation (complete models only)\\
		\hline
	\end{tabular}
\end{table}

\subsection{Atomic diffusion}

Diffusion in {\texttt{MESA}} is based on the Burgers equations \citep{Burgers69}, with the approach of \cite{Thoul94} including the modifications of \cite{Hu11} to include radiative levitation as an option. The normal flag for diffusion in {\texttt{MESA}} only considers thermal and concentration diffusion and gravitational settling, with the inclusion of radiative levitation being included with an additional flag. 

Three separate sets of simulations were carried out. As listed in Table \ref{tab:parameters}, these were, one in which no atomic diffusion was carried out (basic models), one in which {\texttt{MESA}}'s standard atomic diffusion was included (standard models) and a third set of models which included both standard diffusion and radiative levitation (complete models). {\texttt{MESA}} includes gravitational settling and thermal diffusion and concentration diffusion as the standard processes with the {\texttt{use\_element\_diffusion}}
flag, whereas radiative levitation is included as an extra option. The calculation of radiative forces on all of the ions at each mesh point is a computationally expensive task so only a small set of models was produced which included radiative levitation, to investigate the effect it has on the results compared with the standard diffusion experiment. In this work, the elements for which diffusion velocities were calculated are H, He, C, N, O, Ne, Mg, Ar, Cr, Fe and Ni. For the models with radiative levitation, this was only enabled after the onset of helium burning at the first helium flash. Doing so before this caused numerical issues when the model encountered the large convective region driven by the flash. The convection generated would wash away the effects of any diffusion that occurred prior to this in any case. Results for the basic models are presented in Section \ref{sec:basicres}, with the results of the standard and complete simulations in Sections \ref{sec:stdres} and \ref{sec:radlevres} respectively. Diffusion velocities are calculated throughout the entire star for standard diffusion. For the complete models, the standard diffusion component of the calculations (thermal diffusion, concentration diffusion and gravitational settling) is carried out throughout the entire star while diffusion velocities owing to radiative levitation are calculated only in the outer layers of the star with a temperature less than $10^7\,$K. Radiative levitation is only activated after the first helium flash to avoid numerical instabilities.

\subsection{Generating Subdwarf Models}

\begin{figure}
\centering
	\includegraphics[width=0.95\columnwidth]{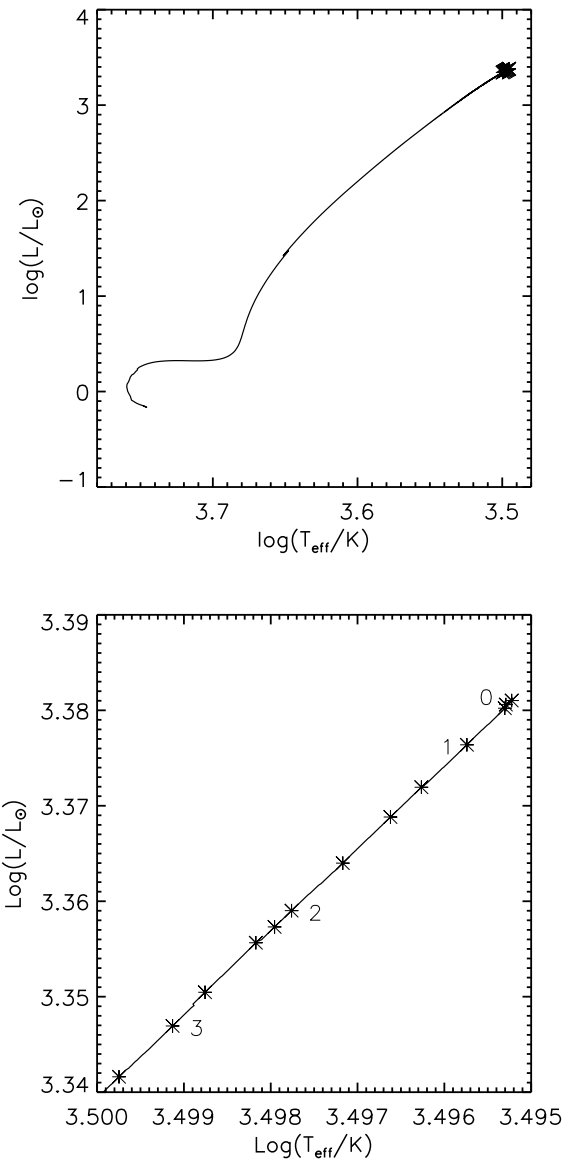}
    \caption{A Hertzsprung-Russell diagram illustrating the evolution of a $1\,M_{\odot}$ star along the RGB. The crosses illustrate the different starting points on the RGB at which common envelope ejection events were carried out, close to the RGB tip. The lower panel shows an expanded view of these points. Three models labelled 0--3 are referred to in the text and subsequent figures.}
    \label{fig:RGB_models}
\end{figure}

We adopt, as subdwarf progenitor models red giant models at various points close to the tip of the red giant branch, shortly before helium core ignition. These models were produced from a 1\,$M_{\odot}$ main sequence (MS) star. No mass loss was included in this phase of evolution because there is little difference in core mass across the range of MS masses which produce degenerate helium ignition ($0.8 - 2.0$\,$M_\odot$) and the fully convective red giant envelope means that the exact mass of the star has little effect on the structure of the stellar envelope. The model was then evolved to the tip of the RGB. At this point, the subdwarf progenitor models were chosen. These models then serve as the starting points for the common envelope phase.  The initial positions of subdwarf progenitor models on the red giant branch are shown in Fig.~\ref{fig:RGB_models}, with the tip of the RGB also indicated. The numbers on the plot indicate specific models at different distances from the RGB tip, with the position of the resulting models on the horizontal branch indicated in subsequent figures. 

The common envelope ejection was approximated by invoking a high mass-loss rate in {\texttt{MESA}} (the {\texttt{relax\_mass}} option which removes mass at a rate of about $10^{-3} M_{\odot}\,\mathrm{yr}^{-1}$) down to a specified envelope mass. In the case of this experiment, the residual envelope mass was $6\times10^{-3}\,M_{\odot}$. This choice of envelope mass was arbitrary, but represented a typical value for a hot subdwarf. Choosing a larger or smaller initial envelope mass for a fixed core mass will shift the zero-age horizontal branch model to cooler or hotter temperatures respectively. Following this rapid mass loss, evolution was followed up to the point of helium core ignition. The simulation was stopped once the central helium mass fraction drops below $0.925$. At this point the models were deemed to have reached the zero-age extended horizontal branch.

\section{Results}

\begin{figure}
\centering
	\includegraphics[width=0.95\columnwidth]{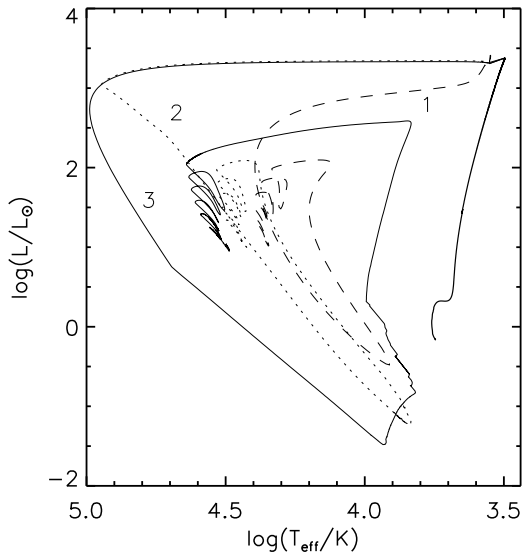}
    \caption{A Hertzsprung-Russell diagram illustrating the evolution of three different post-CEE models. The solid line (1) is that of a very late hot flasher which forms a hot, helium-rich hot subdwarf, the dashed line (3) is an early late hot flasher which forms a cooler, hydrogen rich hot subdwarf, and the dotted line (2) is a model which flashes at an intermediate time after the CEE and forms a hydrogen-rich, intermediate temperature hot subdwarf.}
    \label{fig:Evolution}
\end{figure}

When the star leaves the RGB, hydrogen-shell burning continues, until the core of the star become hot enough and massive enough for helium to ignite. This generally happens slightly off-centre before undergoing a series of smaller flashes, until helium ignition reaches the centre of the star. This explains the `loops' in the evolution tracks shown in Fig. \ref{fig:Evolution}. After helium ignition, the core expands and hydrogen-shell burning is quenched. 

\subsection{Evolution from the RGB to the ZAEHB}

Three different evolution tracks are shown, representing the different stages at which the flash can occur. The numbers next to the tracks indicate which of the numbered progenitor models in Fig.~\ref{fig:RGB_models} they are associated with. The solid line (model 3) shows a model that has a very late flash, owing to the fact that the envelope is removed when it still requires a significant amount of growth in core mass, and has reached the white dwarf cooling track before the core contracts and heats enough to ignite. 
The dashed line (model 1) is the evolution track of a model which underwent a CEE immediately before the point of the first helium flash. No additional burning is needed to reach the threshold mass and the envelope left behind after the ejection is mostly unaltered by the ignition of helium, meaning it arrives on the cooler end of the EHB. The dotted track (model 2) is a model which is intermediate between these two extremes. These evolution tracks are similar to those in the literature \cite[panels c, d and e of their fig. 4]{Brown01}.These are referred to as early, intermediate and late hot flashers, based on how soon after leaving the red giant branch that the first helium flash occurs. The formation of He-rich stars from very late helium flashes is also in agreement with the results of \cite{MillerBertolami08}. The model with the largest core mass (model 0) represents an RGB star at the point of first helium flash. Using sudden mass loss to replicate a common envelope ejection at this point leads to an unphysical result whereby mass loss continues while the star contracts significantly. Thus results for this model have been excluded from subsequent plots.

\begin{figure*}
\centering
	\includegraphics[width=1.9\columnwidth]{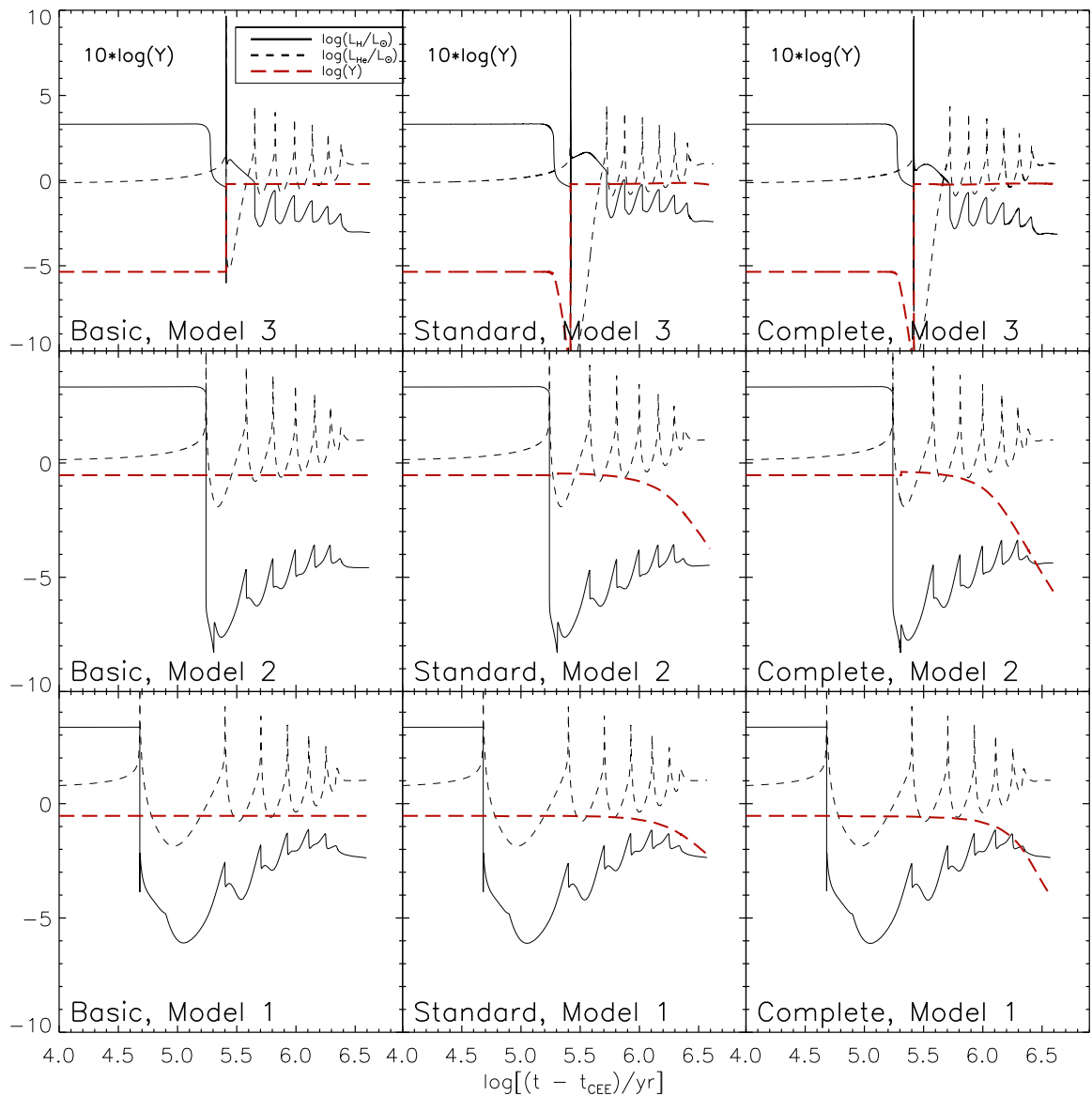}
    \caption{Time evolution of hydrogen luminosity, helium luminosity (both in solar units) and surface helium mass fraction from common envelope ejection up to the point of helium core ignition on the zero-age horizontal branch. The model numbers refer to the representative models indicated in previous figures. The terms basic, standard and complete refers to models with no diffusion, models with diffusion and models with diffusion and radiative levitation respectively. For the top panels, $\log(Y)$ has been multiplied by 10 for clarity.}
    \label{fig:timeevol}
\end{figure*}

The effects of these flashes can be seen in Fig.~\ref{fig:timeevol}, where the time evolution of hydrogen and helium luminosity are plotted, along with surface helium mass fraction. This reveals that model 3, which underwent the common-envelope ejection  with a smaller core mass, takes more time to ignite helium. However when it does, it also has a large hydrogen shell flash which burns some of the residual hydrogen envelope and also drives a significant amount of convective mixing which further depletes the surface of hydrogen, and results in the formation of helium-rich models. Another notable result that can be seen in Fig.~\ref{fig:timeevol} is that, at the point of reaching the zero-age horizontal branch, the surface helium abundance of models which include diffusion is still in a decline and is yet to reach an equilibrium value. Therefore the expected values of the surface helium mass fraction, $\log(Y)$, for hot subdwarfs which are more evolved will be lower than the abundances found in these simulations. Further analysis of an individual example is carried out in Section~\ref{sec:beyond}.

\subsection{Basic models}
\label{sec:basicres}
\begin{figure}
\centering
	\includegraphics[width=0.95\columnwidth]{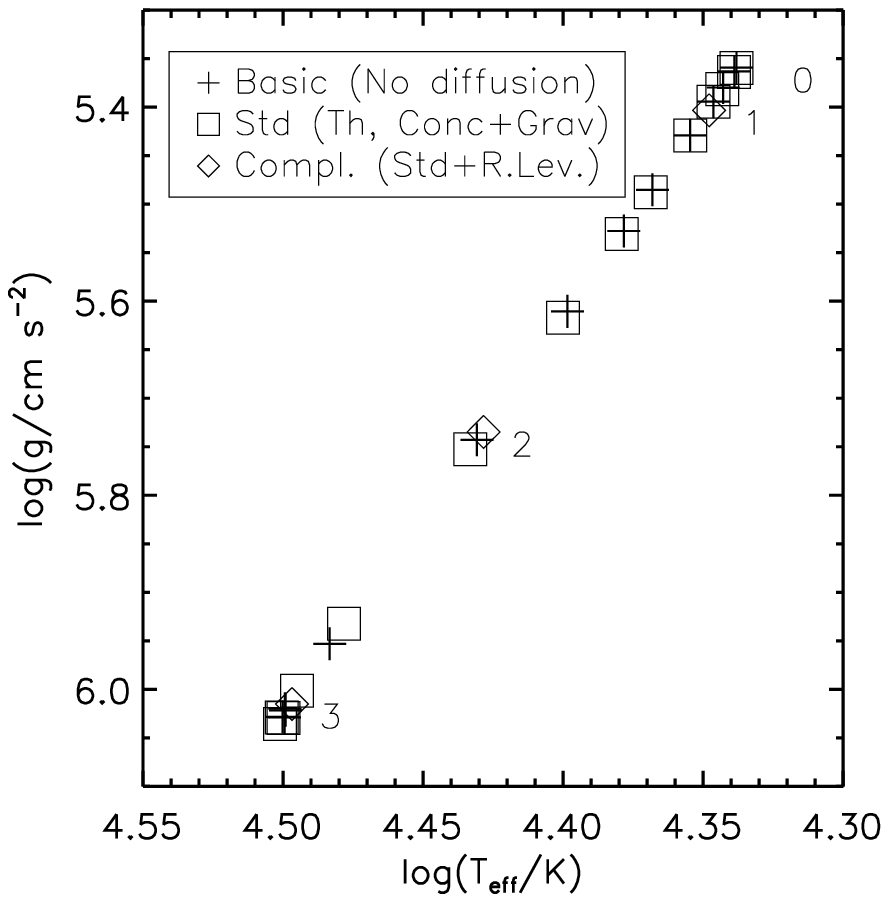}
    \caption{A surface gravity - effective temperature diagram illustrating the zero-age horizontal branch positions of the post-common envelope models.  The three sets of models (basic, standard and complete) are indicated by crosses ($+$), squares ($\square$), and diamonds ($\Diamond$) respectively. The labels 0-3 indicate the approximate positions of the models referred to in the text and in Table~\ref{tab:models}.}
    \label{fig:HB_models}
\end{figure}

A total of 34 models were produced to populate the EHB, using three different sets of diffusion physics. The ZAEHB positions of the models are shown in a surface gravity - effective temperature diagram in Fig. \ref{fig:HB_models}. Diffusion makes very little difference to the ZAEHB positions, as they all populate the same linear region of the gravity-temperature domain. 

Table~\ref{tab:models} shows the key properties of the progenitor and subdwarf models for all 3 sets of input physics, including core mass at departure from the RGB, $\mathrm{M}_{\mathrm{Core}}$,  total subdwarf mass, $\mathrm{M}_{\mathrm{sdB}}$, subdwarf envelope mass, $\mathrm{M}_{\mathrm{Env}}$ and surface helium mass fraction, $\log(Y)$, for zero-age horizontal branch models for the different diffusion options. The mass contained between the surface and the point in the star where the hydrogen mass fraction drops below $0.01$ gives the value of $\mathrm{M}_{\mathrm{Env}}$. $\log(Y)$ and all subsequent elemental mass fractions refer to the average abundance of the outer $10^{-8}$ of the total stellar mass. The final envelope mass reported in Table~\ref{tab:models} differs from the initial envelope mass for a number of reasons. First, the core mass at envelope ejection is a key quantity. Models further from the tip of the giant branch need to grow larger cores before igniting helium and thus consume some of the remaining envelope before becoming a horizontal branch star. In the case of the very late flashers, the envelopes are so thin at the time of helium flash that the flash driven convection leads to the formation of a helium-rich star. Secondly, the choice of diffusion physics affects the reported envelope mass. This is due to the settling of helium in the models including diffusion, which shifts the position where the hydrogen mass fraction drops below 0.01 deeper into the star. Hence some of the models close to the tip of the red giant branch end up with zero-age horizontal branch envelope masses slightly larger than their initial values.

\begin{table*} 

\centering
	\caption{Summary of key properties of the subdwarf models produced and the core mass of their corresponding progenitor. The numbered models referred to in figures and the text are also indicated.}
	\label{tab:models}
	\begin{tabular}{ccc|cc|cc|cc}
		\hline
		\multicolumn{3}{c}{ }&\multicolumn{2}{c}{Basic (No Diffusion)}&\multicolumn{2}{c}{Standard Diffusion}&\multicolumn{2}{c}{Complete Diffusion}\\ \hline
N$_{\mathrm{model}}$ & $\mathrm{M}_{\mathrm{Core}}/\mathrm{M}_\odot$ & $\mathrm{M}_{\mathrm{sdB}}/\mathrm{M}_\odot$ & $\mathrm{M}_{\mathrm{Env}}/\mathrm{M}_\odot$ & $\log(Y)$& $\mathrm{M}_{\mathrm{Env}}/\mathrm{M}_\odot$ & $\log(Y)$ & $\mathrm{M}_{\mathrm{Env}}/\mathrm{M}_\odot$ & $\log(Y)$ \\ \hline
--	&	0.4550   	& 0.4610   	& --   						& -0.0204   & 9.366$\times10^{-9}$		& -0.0172 	& -- 					& -- 		\\
3	&	0.4561   	& 0.4621   	& --   						& -0.0203   & 1.488$\times10^{-8}$  	& -0.0291 	& 1.156$\times10^{-9}$	& -0.0231 	\\
--	&	0.4569   	& 0.4629   	& --   						& -0.0218	& 1.620$\times10^{-8}$   	& -0.0336 	& -- 					& -- 		\\
--	&	0.4581   	& 0.4641   	& --   						& -0.0220 	& 2.633$\times10^{-6}$   	& -1.8928 	& -- 					& -- 		\\
--	&	0.4585   	& 0.4645   	& 4.607$\times10^{-3}$   	& -0.0717 	& 5.161$\times10^{-3}$    	& -4.8077 	& -- 					& -- 		\\
2	&	0.4588   	& 0.4648   	& 2.254$\times10^{-3}$   	& -0.5356	& 2.286$\times10^{-3}$   	& -3.7528 	& 2.392$\times10^{-3}$  & -5.6146 	\\
--	&	0.4600   	& 0.4660   	& 1.605$\times10^{-3}$   	& -0.5356	& 1.853$\times10^{-3}$  	& -3.0379 	& -- 					& -- 		\\
--	&	0.4611   	& 0.4671   	& 2.676$\times10^{-3}$   	& -0.5356	& 3.052$\times10^{-3}$   	& -2.6275 	& -- 					& -- 		\\
--	&	0.4619   	& 0.4679   	& 3.374$\times10^{-3}$   	& -0.5356	& 3.838$\times10^{-3}$   	& -2.4416 	& -- 					& -- 		\\
1	&	0.4630   	& 0.4690   	& 4.450$\times10^{-3}$   	& -0.5356	& 5.021$\times10^{-3}$   	& -2.2191 	& 5.006$\times10^{-3}$ 	& -4.0330 	\\
--	&	0.4637   	& 0.4698   	& 5.147$\times10^{-3}$   	& -0.5356	& 5.805$\times10^{-3}$   	& -2.0921 	& -- 					& -- 		\\
--	&	0.4641   	& 0.4701   	& 5.498$\times10^{-3}$   	& -0.5356	& 6.144$\times10^{-3}$   	& -2.0489 	& -- 					& -- 		\\
--	&	0.4645   	& 0.4705   	& 5.853$\times10^{-3}$   	& -0.5356	& 6.540$\times10^{-3}$   	& -1.9803 	& -- 					& -- 		\\
--	&	0.4646 		& 0.4706   	& 6.003$\times10^{-3}$   	& -0.5356	& 6.687$\times10^{-3}$   	& -1.9786 	& -- 					& -- 		\\
0	&	0.4647 		& 0.4707   	& 6.014$\times10^{-3}$   	& -0.5356	& 6.790$\times10^{-3}$   	& -2.3461	& 6.797$\times10^{-3}$  & -4.3458	\\
		\hline
	\end{tabular}
\end{table*}

These models have a spread of effective temperatures in the range of $22\,000 - 32\,000\,\mathrm{K}$ ($4.34\le \log T_{\mathrm{eff}}/\mathrm{K}\le 4.51$), where the surface temperature is directly related to the mass of hydrogen envelope remaining. This variation in hydrogen envelope mass is partially due to the time taken for stars to ignite helium. The models which undergo a CEE event earlier on the RGB start with a smaller core, and spend time burning some of the remaining envelope hydrogen into helium to achieve a large enough core for helium burning to begin. 

\begin{figure}
\centering
	\includegraphics[width=0.95\columnwidth]{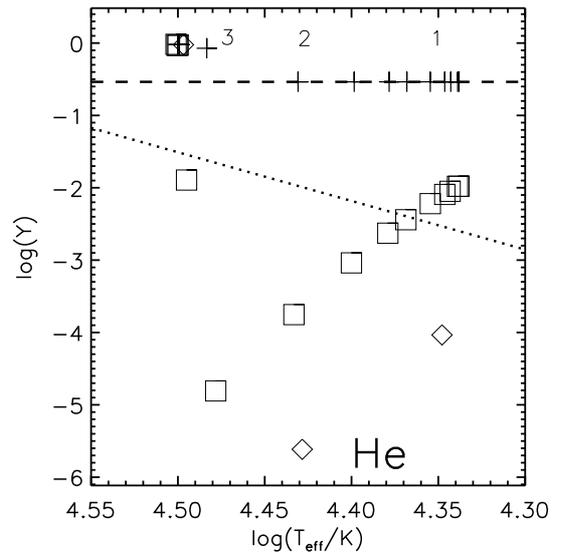}
    \caption{Zero-age horizontal branch surface helium abundances (mass fraction) as a function of effective temperature. The symbols have the same meaning as in Fig. \ref{fig:HB_models}. The numerals indicate the basic models corresponding to the representative progenitor models also numbered in Fig.~\ref{fig:Evolution}. The dashed and dotted lines indicate solar values and observational values for hot subdwarfs from the work of \protect\cite{Geier13} respectively. The observational values are based on an interpolation between average `cool' and `warm' sdB abundances as in Table 2 of \protect\cite{Jeffery17}.}
    \label{fig:heteff}
\end{figure}

Surface helium abundances were calculated for the zero-age horizontal branch models and were plotted as a function of the effective temperature. The results of this can be seen in Fig. \ref{fig:heteff}, with the basic models being indicated by the crosses (+). The solar helium mass fraction is  indicated by the dashed line, while the dotted line represents typical sdB mass fractions found by \cite{Geier13}. These typical observational abundances are presented here as a linear interpolation between the average abundances of `cool' and `warm' sdBs as in Table 2 of \cite{Jeffery17}.

This behaviour is quite similar to that seen in other theoretical models of hot subdwarfs produced with {\texttt{MESA}} where the hottest models tend to form helium-rich subdwarfs due to convective mixing of the remaining low-mass envelope, as seen in fig. 9 of \cite{Xiong17}, for example. The distribution of abundances and temperatures is far from a good match to observations. This will be discussed in more detail in section~\ref{compobs}.

\subsection{Standard diffusion models}
\label{sec:stdres}
To investigate the effects of atomic diffusion (with the exception of radiative levitation) on the evolution of these stars, a second set of models was produced via identical methods, with diffusion included. The luminosity and effective temperatures of the models were mostly unchanged, as shown in Fig. \ref{fig:HB_models}.

The effects of diffusion are much more noticeable in the surface helium mass fractions. These models show a significant depletion of helium in the envelopes of the hydrogen-rich subdwarfs, which increases with effective temperature. This is a consequence of the surface gravity being higher in the hotter stars. As soon as the hydrogen envelope has become thin enough for it to be mixed during the helium shell flashes, the decline in helium abundance is reversed and the models develop helium-dominated atmospheres. From this it can be inferred that, in order to form a hydrogen-rich hot subdwarf on the zero-age horizontal branch, a red giant must be quite close to undergoing a helium flash at the time of common-envelope ejection. However, helium-deficient subdwarfs are known to exist at temperatures higher than our models predict, so the choice of input physics in these simulations may not accurately represent the aftermath of a common-envelope event. For example, as suggested by Fig.~\ref{fig:timeevol}, and expanded upon in Section~\ref{sec:beyond} the surface helium abundance generally declines over the horizontal branch lifetime owing to diffusion, therefore such stars are likely more evolved stars. Additionally, over the horizontal branch lifetime, hot subdwarfs move to hotter temperatures and higher luminosities.

\subsection{Complete diffusion models}
\label{sec:radlevres}

Radiative levitation has been known to play an important role in hot subdwarf physics. Levitation of iron has been shown to cause an opacity bump large enough to produce the pulsations observed in many subdwarfs \citep{Charpinet97}. Later studies showed that nickel opacity also played a key role in the driving of these pulsations \citep{JefferySaio07}. Due to the computational effort required to determine radiative forces on all ions at all depths within the model stars, a smaller set of models (4 in total) was produced in order to determine the change in results expected from the inclusion of all aspects of atomic diffusion at different positions along the EHB. The diamonds in Fig.~\ref{fig:heteff} show that radiative levitation leads to further depletion of helium from the atmosphere in the pre-horizontal branch phase, for the cooler, hydrogen rich models, while the helium abundance of the helium-rich model is comparable to that of the standard diffusion model.
Experiments with diffusion only being calculated for H and He show a similar difference between the models with and without radiative levitation, regardless of the inclusion of other elements in the calculations. Some modelling of the evolution on the extreme horizontal branch was carried out to investigate this further and is presented in Section~\ref{sec:beyond}. This can also be seen in the time evolution of $\log(Y)$ in Fig.~\ref{fig:timeevol}, where the complete diffusion models show a more pronounced decline in surface helium than the corresponding standard models. 

\subsection{Beyond the Zero-Age Horizontal Branch}
\label{sec:beyond}
\begin{figure}
\includegraphics[width=0.45\textwidth]{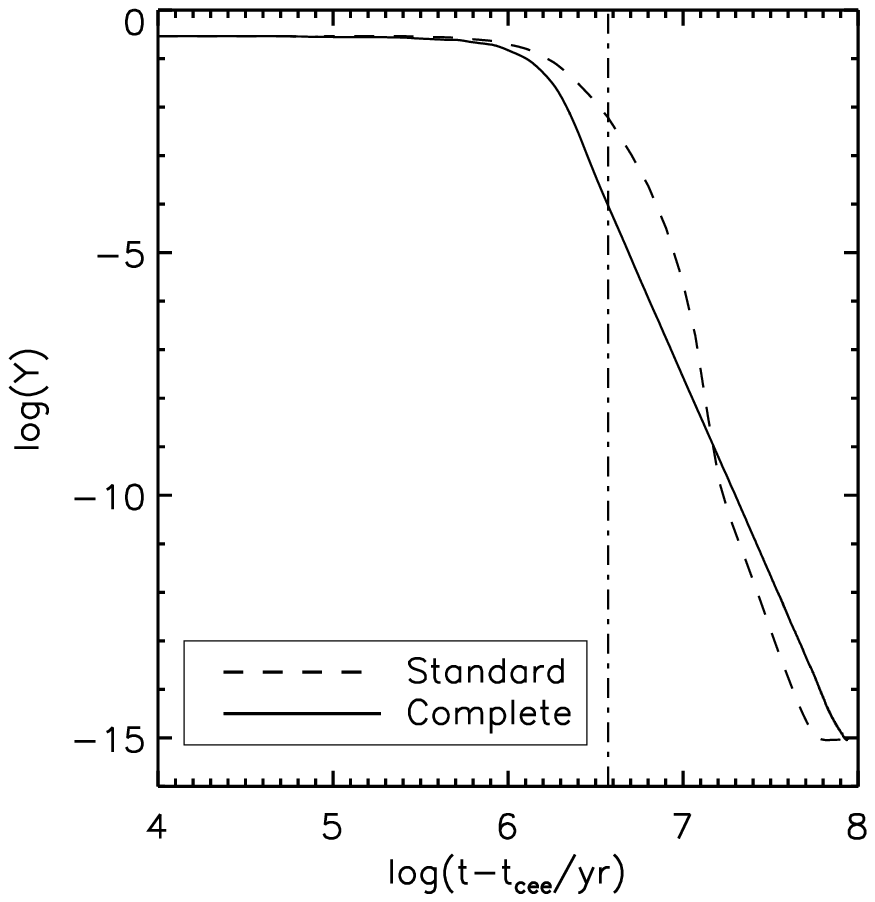}
\includegraphics[width=0.45\textwidth]{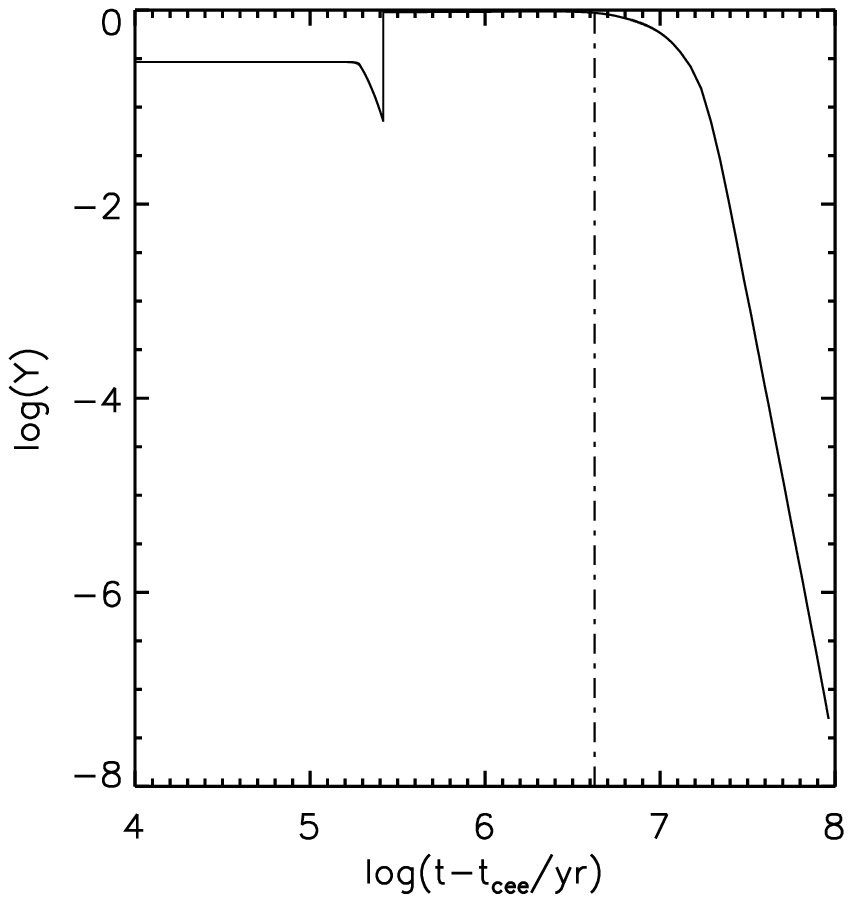}
\caption{\emph{Top:} Surface helium mass fraction as a function of time for Model 1 from post-common envelope phase right through to the end of core helium burning, with the dashed line representing a model with standard diffusion, while the complete diffusion model is shown by the solid line. The vertical dot-dashed line indicates the time at which the models reached the zero-age horizontal branch. \emph{Bottom:} Surface helium mass fraction as a function of time for Model 3 with standard diffusion. The vertical dot-dashed line indicates the time at which the model reached the zero-age horizontal branch.}
\label{fig:beyond}
\end{figure}

The presence of a difference between surface helium abundances in the zero-age horizontal branch models with and without radiative levitation is an interesting result. In order to investigate further, a selected model (model 1) was taken from the post-common envelope stage all the way to the end of the horizontal branch (determined as the point where central helium mass fraction reaches $0.1$). Both the standard and complete diffusion options were simulated. The evolution of $\log(Y)$ is shown in the upper panel of Fig.~\ref{fig:beyond}. Here it can be seen that discrepancy continues for most of the horizontal branch lifetime, reaching an equilibrium at an insignificant level of helium towards the end of the horizontal branch. The vertical line illustrates the time at which the zero-age horizontal branch is reached, and illustrates the difference seen between the results shown in Fig.~\ref{fig:heteff}. The fact that both models show an insignificant amount of helium by the end of their lifetime suggests that mass loss and/or other processes must reduce the effectiveness of diffusion in order to match observed helium abundances.

The other issue which onward evolution can address is the fate of the helium-rich zero-age models present in these results. Using standard diffusion, model 3 was evolved along the horizontal branch. The evolution of $\log(Y)$ is shown in the bottom panel of Fig.~\ref{fig:beyond}. The helium enrichment of the surface due to the hydrogen shell flash is quite evident. The model remains helium rich at the zero-age horizontal branch (again shown by a vertical line). Beyond the zero-age horizontal branch however, the remaining hydrogen begins to resurface and $\log(Y)$ decreases over the horizontal branch lifetime. This indicates that helium-rich and intermediate helium-rich subdwarfs are more likely to be young extreme horizontal branch stars or even pre-horizontal branch objects.

\subsection{Differences between standard and complete diffusion models}
\begin{figure}
\centering
\includegraphics[width=0.38\textwidth]{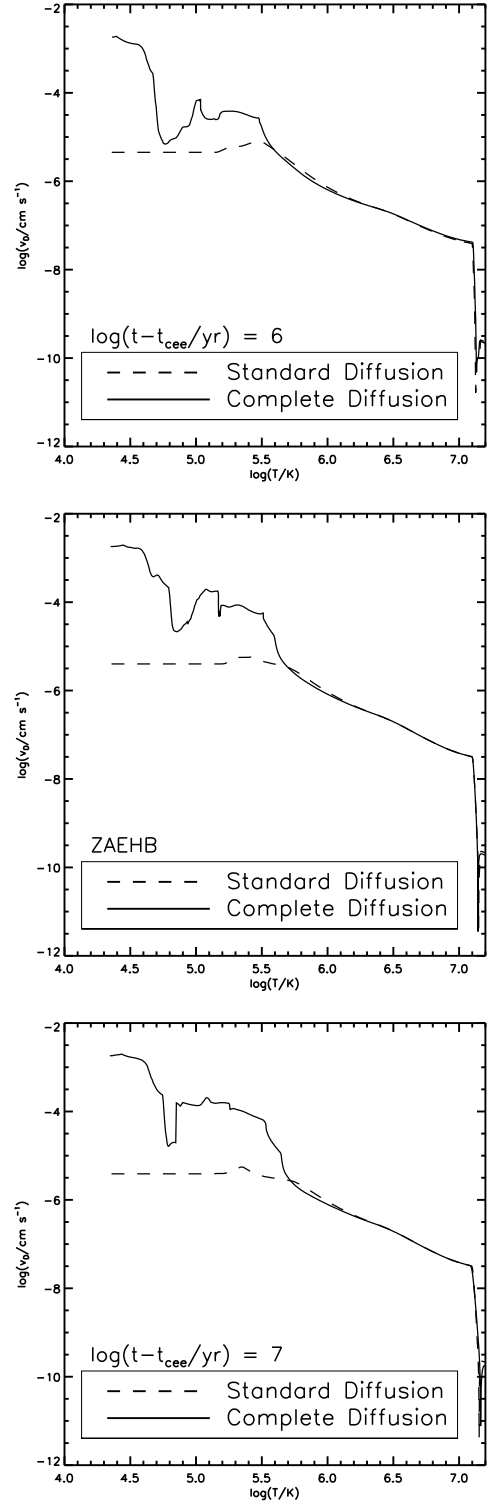}
\caption{Diffusion velocity ($v_D$) of helium as a function of temperature in model 1 at 3 separate stages of evolution, $10^6$ years after envelope ejection (top panel), at the zero age horizontal branch (approximately $10^{6.6}$ years after envelope ejection, middle panel) and $10^7$ years after envelope ejection. The results for the standard and complete model are shown by the dashed and solid lines respectively. The value of $v_D$ is negative up to temperatures of $10^{7.1}\,$K where a sign inversion takes place.}
\label{fig:vdiff}
\end{figure}

The behaviour of helium following the introduction of radiative levitation is interesting. Fig~\ref{fig:vdiff} illustrates the diffusion velocity of helium in model 1 as a function of temperature. In all cases, the outer layers of the model have a much larger negative diffusion velocity when radiative levitation is included. This larger velocity leads to a larger  reduction in the surface helium abundance. Once the mass fraction of a particular element in a cell drops below $10^{-15}$, a diffusion velocity is no longer calculated. Once the helium mass fraction of all the outermost cells drops below this diffusion of helium is no longer calculated, producing the flat line at lower-right in the top panel of Fig.\ref{fig:beyond}. The other noticeable feature in Fig.~\ref{fig:vdiff} is the constant diffusion velocity for temperatures $\log T/{\rm K}<5.2$ in the standard models. This results from the \texttt{MESA} input parameter choice for standard diffusion, which consequently treats a number of the outermost cells as a single entity to improve numerical stability. This approximation alone is not responsible for the velocity difference, because the radiative levitation velocities become substantially greater than the standard diffusion velocities at $5.2<\log T/{\rm K}<5.5$. Reduction of cell sizes in the standard diffusion calculation is not expected to lead to any significant difference in the results, and the difference between the surface abundance of helium in the standard and complete models would persist.

\section{Discussion}
\label{sec:discussion}

\subsection{Comparison to other evolutionary models}
The set of basic models reproduce quite well the results of \cite{Xiong17}, both in terms of surface abundances and distribution on the HR diagram. The stars cooler than $28\,000\,\mathrm{K}$ ($\log T_{\mathrm{eff}} = 4.45$) retain the chemical composition of the envelope that they had while on the RGB and immediately after the common envelope ejection. The stars hotter than $30\,000\,\mathrm{K}$ ($\log T_{\mathrm{eff}} = 4.48$) all show significant helium enhancement, while a gap in temperature distribution is seen between the two groups of stars. This is a consequence of having envelopes sufficiently small. In this case, hydrogen gets depleted by the hydrogen shell flash and associated convective mixing of the core and envelope. Rather than a uniform decrease in envelope size, there is thus a point at which the hydrogen envelope is too thin to survive the flash-driven mixing and these models develop helium-rich atmospheres. This leads to a jump in the effective temperature distribution also, due to the sudden change in the mean atomic weight of the envelope.

The inclusion of diffusion makes a substantial difference to the surface composition of the models. Gravitational settling is more dominant in the hotter stars, which have a higher surface gravity, thus hydrogen becomes more dominant at higher temperatures. This trend gets reversed once the flash mixing is able to penetrate through the entire envelope. Helium becomes dominant from this point onwards for the zero age models. Recovery of any remaining hydrogen to the surface is likely over the horizontal branch lifetime. The addition of radiative levitation further depletes the helium from the hydrogen-rich atmospheres. The helium-rich model (model 3) with radiative levitation has a similar helium abundance to that of the corresponding `standard' diffusion model. This is in stark contrast to observations which show a general trend for the helium abundance to increase with increasing temperature.

\subsection{Comparison to other diffusion studies}
\label{disc:diff}

This work also set out to investigate the effects of diffusion on the surface abundances of other elements. Studies into the effects of diffusion on hot subdwarf stars have been carried out before (\cite{Michaud11, Hu11} for example). However these studies used approximate methods for the helium flash in order to generate their subdwarf models. This study takes advantage of the fact that {\texttt{MESA}} is capable of evolving through the helium flash, albeit still with a certain amount of approximation. Using {\texttt{MESA}}, however, allows the effects of diffusion in the evolution from common envelope ejection through to core helium ignition to be examined in a more self-consistent manner. It is also worth noting that this study does still require the use of approximations for the common-envelope phase. By evolving through the helium flash, it was possible to investigate the effects of diffusion in the pre-horizontal branch phase of evolution. Under the assumption that the star is stable enough for atomic diffusion, it is found that diffusion plays a role in the surface properties of these stars before they reach the EHB, compared to earlier studies which start with a ZAEHB model for their diffusion studies.

\begin{figure*}
\centering

	\includegraphics[width=0.6\columnwidth]{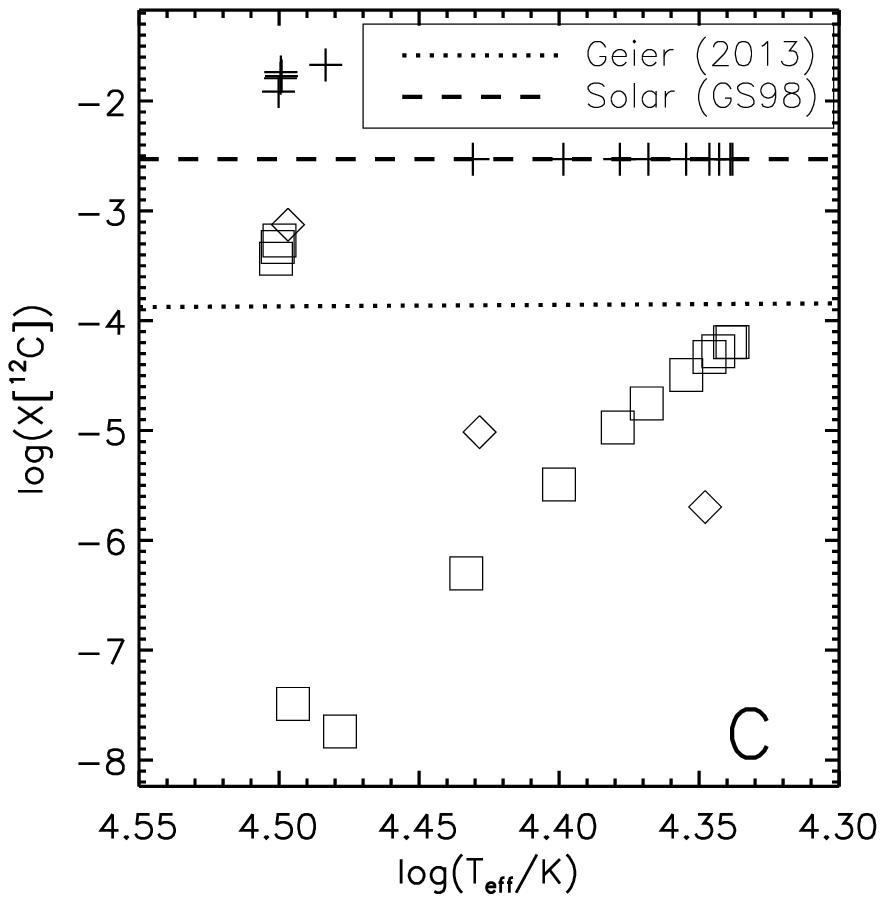}
	\includegraphics[width=0.6\columnwidth]{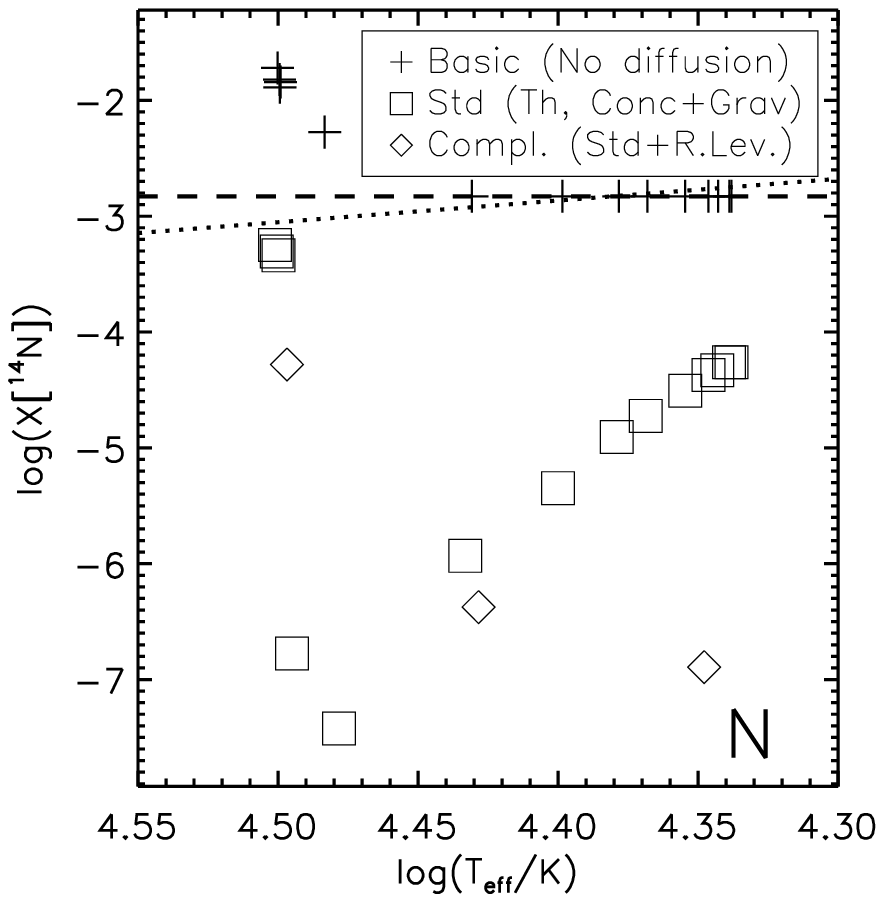}
	\includegraphics[width=0.6\columnwidth]{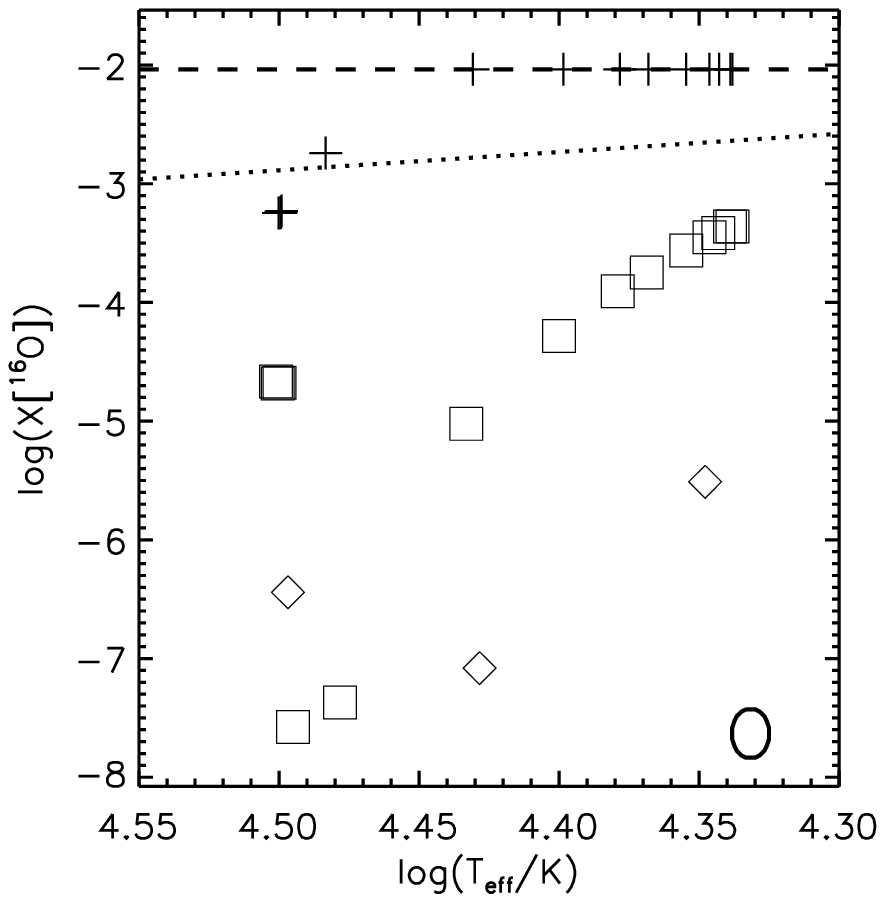}
	\includegraphics[width=0.6\columnwidth]{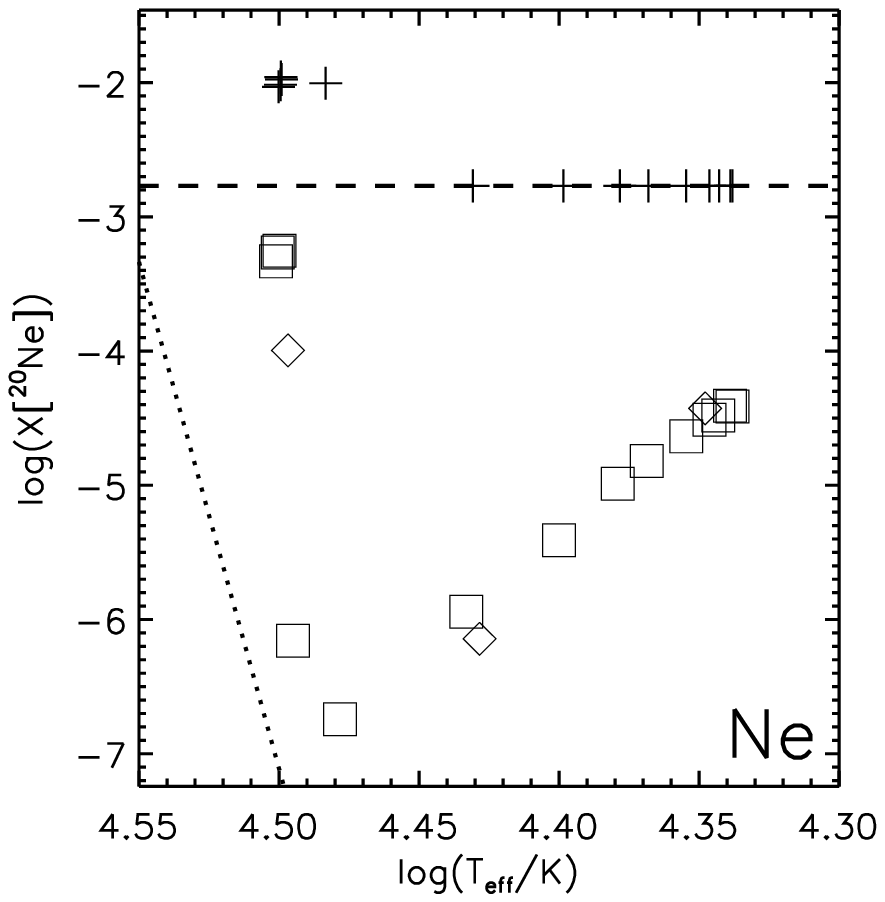}
	\includegraphics[width=0.6\columnwidth]{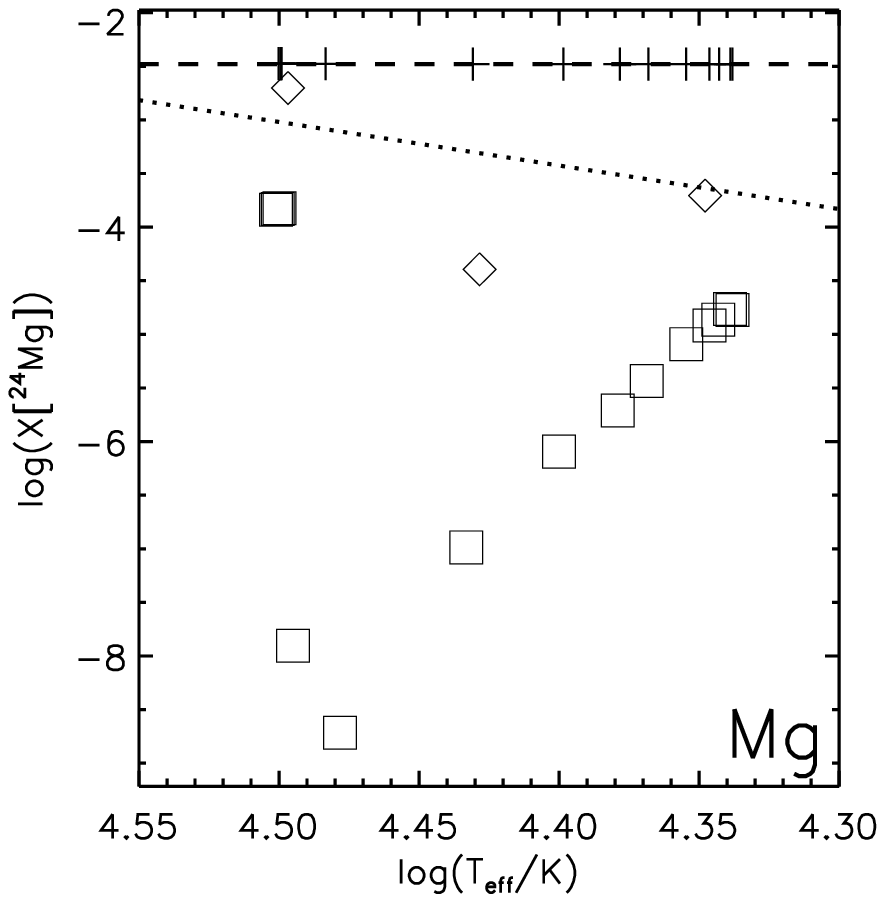}
	\includegraphics[width=0.6\columnwidth]{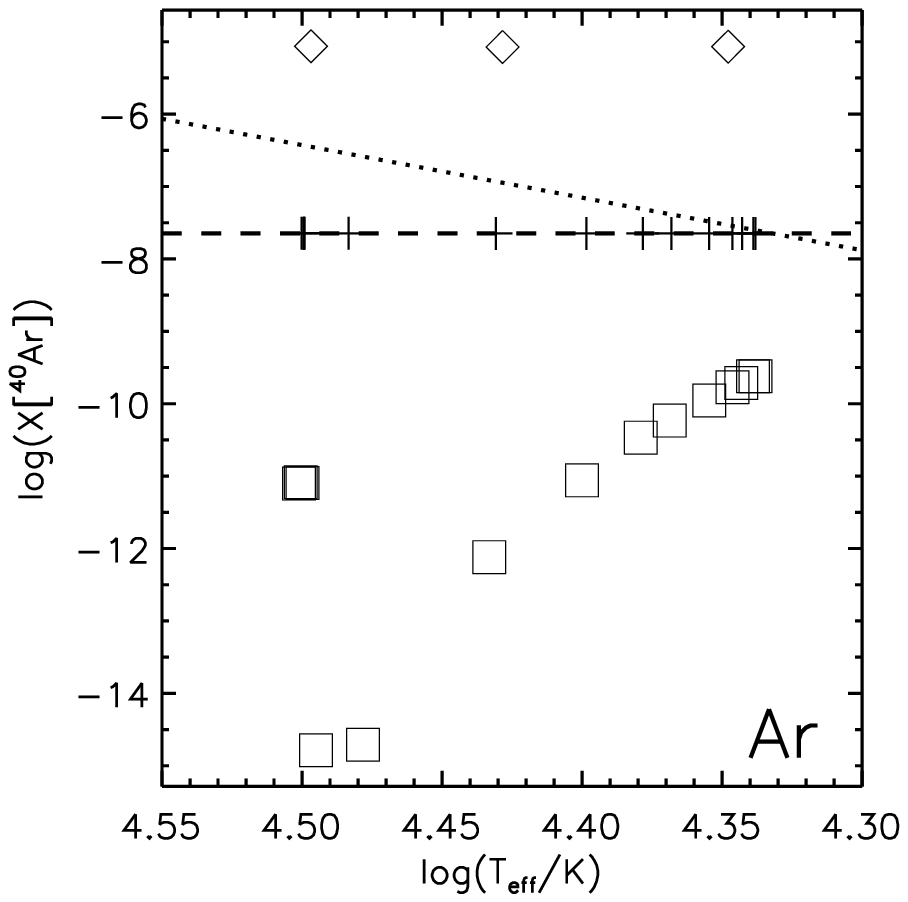}
	\includegraphics[width=0.6\columnwidth]{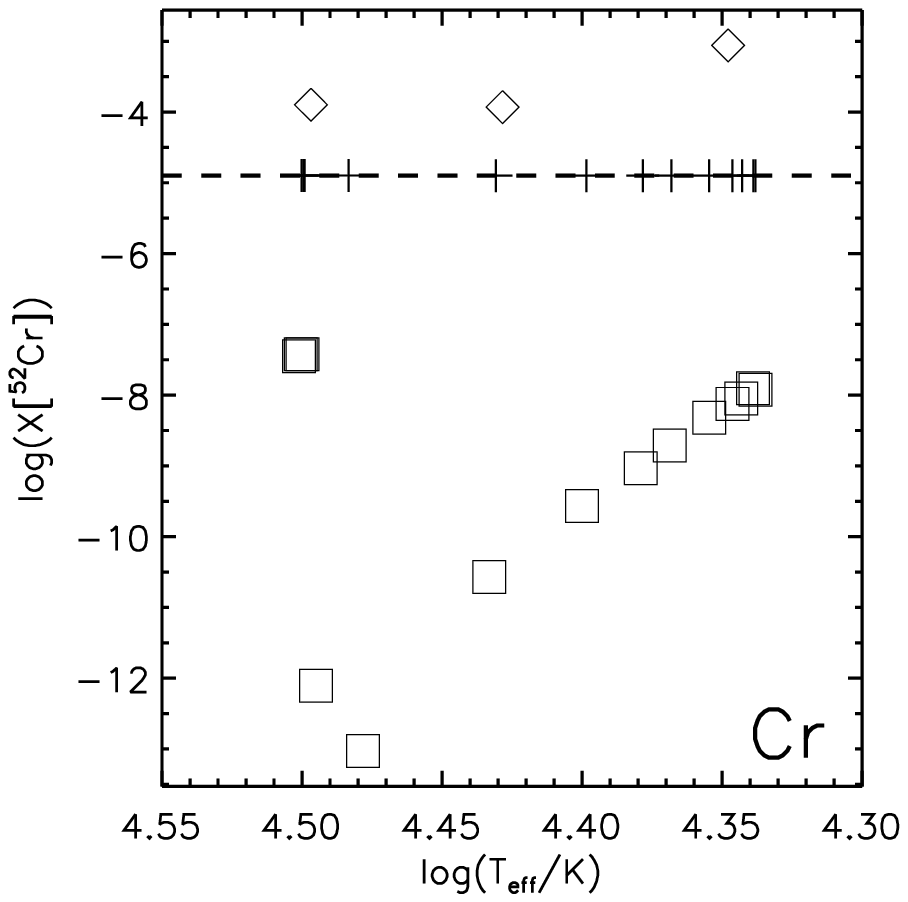}
	\includegraphics[width=0.6\columnwidth]{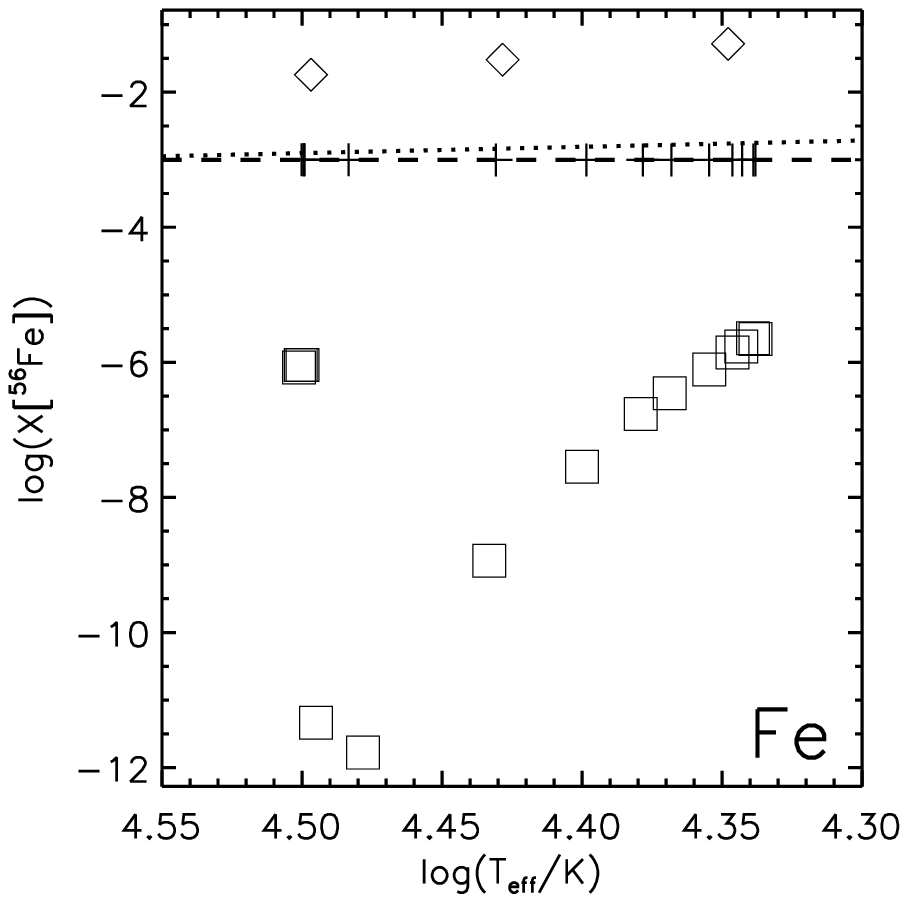}
	\includegraphics[width=0.6\columnwidth]{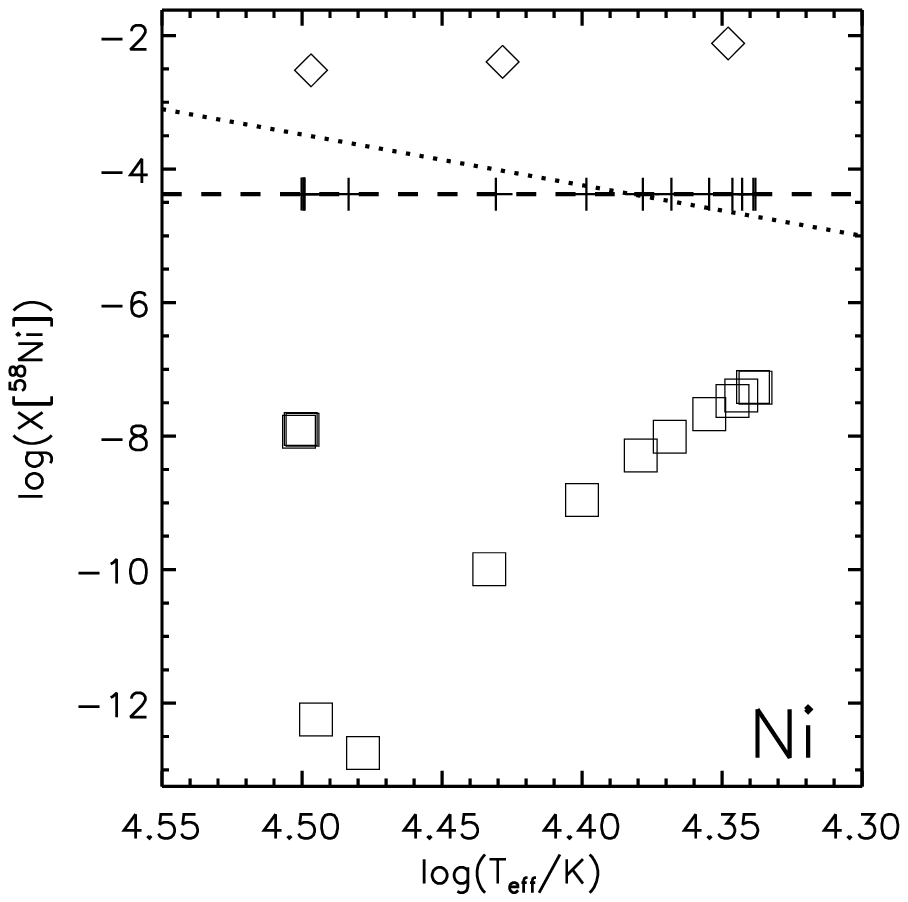}
    \caption{Zero-age horizontal branch surface abundances (mass fractions) as a function of temperature for the elements included in the models produced in this work. The symbols have the same meaning as in Fig.~\ref{fig:HB_models}. The dashed and the dotted lines have the same meaning as in Fig.~\ref{fig:heteff}}
    \label{fig:elementeff}
\end{figure*}

Fig. \ref{fig:elementeff} shows the surface abundances of the elements which were included in the diffusion calculations, plotted as a function of the ZAEHB temperatures of the models. The symbols have the same meaning as in the previous plots. The abundances are shown as mass fractions of the surface composition. With the basic models with no diffusion, most elements maintain their initial abundances across the entire set of models. The only exceptions to this are carbon, nitrogen and neon, which become enriched and oxygen which becomes depleted. This is a direct consequence of the CNO and $\alpha$ processes at work during the hydrogen shell flash. In the standard models, the behaviour of the cool, hydrogen-rich models is comparable to that of helium, with all elements increasingly depleted with increasing surface gravity. In the flash-mixed models, the mass fractions show an increase. However, they remain below the initial values found in the models with no diffusion.

The behaviour of elemental surface abundances as a function of temperature upon the inclusion of radiative levitation is markedly different. Ar, Cr, Fe and Ni all show significant overabundances. This is expected, particularly for Fe and Ni, where the enhancement due to radiative levitation is known to explain the presence of pulsations in some hot subdwarfs \citep{Charpinet97,Fontaine03,JefferySaio06b} and also reasonable given the complex atomic spectra of these elements. Radiative levitation seems to have no effect on Ne, with the data points lying in the same region of the plot as for standard diffusion. The behaviour of N and O is similar to He, while the depletion of Mg from the atmosphere is less significant, meaning it is partially supported by radiation pressure.

These results are in broad agreement with the findings of other studies. \cite{Michaud11} shows the depletion of most light elements and the enhancement of elements heavier than argon, with the exception of iron. Our results which include radiative levitation agree qualitatively with these findings, although a moderate enhancement to the iron abundance is also found. It is worth noting that these simulations do not use the turbulent mixing used by \cite{Michaud11} to obtain their results. In the models of \cite{Hu11} which include either mass loss or turbulent mixing, a slight enhancement of iron abundance is also seen, so a small enhancement of iron in the atmosphere is not an unusual result.

\subsection{Comparison to observations}
\label{compobs}

The hottest subdwarfs produced by this method are the helium-rich models at about $32\,000\,\mathrm{K}$, which is slightly cooler than would be expected for subdwarf B stars, which have temperatures of up to $35\,000\,\mathrm{K}$ or $40\,000\,\mathrm{K}$. However, this issue is not unique to these models. Fig. 4 of \cite{Nemeth12} includes the theoretical ZAEHB from the work of \cite{Dorman93} in the surface gravity - effective temperature domain and the results agree extremely well with the ZAEHB produced in this work.  There are a number of ways by which hotter subdwarfs could be produced, such as more massive stars, or allowing the stars to evolve away from the zero-age horizontal branch.

For models cooler than $28\,000\,\mathrm{K}$ that are not flash mixed to the surface, models with diffusion give a surface helium abundance that is too low when compared to the observed data of \cite{Nemeth12}, while the models with no diffusion (like the models of \cite{Xiong17}) give helium abundances which are too large for the vast majority of subdwarfs in this temperature range. Including mass loss and/or turbulent mixing in the outer layers of hot subdwarfs has been shown to improve the abundance mismatch between observations and models \citep{Hu11}. However, this comes at the expense of affecting the ability of the stars to pulsate and a physical explanation for the turbulent mixing remains unclear. The results presented here also imply that some other mechanism must reduce the effectiveness of the diffusion processes in order to produce subdwarfs with the observed surface helium abundances. The lack of observational evidence for a large population of helium-rich subdwarfs in short period binaries suggests that this evolutionary scenario is not favourable. However, resurfacing of hydrogen in some of the flash-mixed models reduces the number of helium-rich subdwarfs expected. Some close binaries containing helium-rich hot subdwarfs are known to exist. One such example is the helium-rich subdwarf CPD--20$^{\circ}\,1123$ \citep{Naslim12}, which is believed to be a post-common-envelope object, given its orbital period of $2.3\,$d. However, the surface temperature of this star has been measured at around $23\,000\,\mathrm{K}$ ($\log \mathrm{T}_{\mathrm{eff}}/\mathrm{K} = 4.36$), much cooler than the temperatures at which the post-common envelope models presented here (either with or without diffusion) begin to develop super-solar helium abundances due to flash-driven convective mixing of the envelope. This disagreement between theory and observation suggests a problem in the assumptions made in the stellar evolution models. This could potentially be the treatment of diffusion physics or the method of simulating the common envelope ejection.

Another area of input physics which may be causing a discrepancy is the treatment of convection in 1-dimensional stellar evolution codes. Studies of convection in evolution models of subdwarf B stars have been carried out, and show that the size of convective cores in the models are considerably smaller than the convective core masses calculated from asteroseismic data \citep{Schindler15}. This implies that the Schwarzchild criterion, which is typically used to determine whether material is convective, may underestimate the true extent of convective mixing. The implications for the results presented here may be that even larger hydrogen envelopes could be fully mixed by a hydrogen shell flash, leading to a larger number of models which are helium-rich on the zero-age horizontal branch.

In terms of the abundances of other metals, the observational results of \cite{Edelmann01} show depletion of C, N, O and Mg, super-solar Ar abundances and roughly solar Fe. For the complete models which include radiative levitation, the behaviour of these elements is in agreement with the observations, although the magnitude of the depletion is much higher than found observationally (2-3 dex below solar, rather than about 1 dex found in the observations). As discussed earlier, this is likely due to the need for an additional mixing process to oppose the action of radiative levitation. Other observational data for metal abundances in hot subdwarfs \citep{Geier13} show similar behaviour with light elements being depleted and heavier elements being enhanced. However, the magnitude of the changes are not as extreme as the results of the simulations which included radiative levitation.
\section{Conclusions}
\label{sec:conclusion}

Three sequences of post-common envelope stellar evolution models, one without diffusion and two with different diffusion physics, were created. These models had a self-consistent evolution from the red giant branch, through the helium flash to the onset of helium core burning at the zero-age horizontal branch. 

The surface abundances of many elements were examined when the models reached the horizontal branch. The most significant result was the formation of helium-rich models at effective temperatures above $30\,000\,$K. This was found to  be due to a large hydrogen shell flash and significant convective mixing. This only happens when the model leaves the red giant branch due to common-envelope ejection, much earlier than the tip of the giant branch. In these cases, much of the remnant of the envelope is burnt before the core grows massive enough to ignite helium, when it has already reached the white dwarf cooling track. This was found to be the case regardless of the diffusion physics used. However, the abundances of other elements were highly dependent on which diffusion processes were included. This helium-rich phase will not necessarily last for the entirety of the horizontal branch lifetime, as any remaining hydrogen in the envelope will diffuse towards the surface.

For the basic models with no diffusion, all elements remained at their initial abundances, apart from the flash-mixed, helium-rich models where carbon, nitrogen and neon were enhanced and oxygen was depleted due to the enrichment of the surface by CNO processed material from the hydrogen shell flash. With diffusion included, all elements are depleted in the hydrogen-rich models, with abundances increased in the flash-mixed models, but less than their initial abundances without diffusion, assuming the star is stable enough for diffusion during this phase of evolution. When radiative levitation is also included, elements lighter than neon become depleted even more than in the standard diffusion scenario, while elements heavier than magnesium are enhanced.

The results suggest that a population of helium-rich subdwarfs in short period binaries should be seen with temperatures of $30\,000 - 32\,000\,$K, produced after common envelope events. However very few such systems are known to exist. This indicates that in order to produce hydrogen-rich subdwarfs at these temperatures, common envelope ejection must happen very close to the tip of the red giant branch, or a mechanism must exist for material ejected in the common envelope event to be accreted back on to the surface of the star. Another consideration is that in general, the observed stars will have evolved away from the zero-age horizontal branch to a varying degree. This will allow any remaining hydrogen to re-surface. The time evolution of the surface helium abundance shows it is still in decline at the zero-age horizontal branch and is yet to reach an equilibrium. Further study of the evolution beyond the zero-age horizontal branch could provide insight into the origins of intermediate helium-rich subdwarfs.

On the other hand, the cooler, hydrogen-rich models have practically no helium which implies that some form of turbulence or additional mixing must be present in order to reduce the effectiveness of diffusion. The results of \cite{Schindler15}, which suggests that convection is presently underestimated in 1-D evolution models may be an additional source of mixing to counteract levitation and provide a closer match to observations.

Comparisons of these results to the observations of the proposed post-common envelope hot subdwarf CPD--20$^{\circ}\,1123$ shows that helium-rich hot subdwarfs are only produced at temperatures greater than $28\,000\,\mathrm{K}$, which is too hot to explain such a system. Given that the effective temperature of CPD--20$^{\circ}\,1123$ is quite well determined, this suggests a flaw in the evolution models. The two key areas of input physics where there is uncertainty in the models is the treatment of convection and the treatment of the common envelope ejection itself.

\section*{Acknowledgements}

CMB acknowledges funding from the Irish Research Council (Grant No. GOIP 2015/1603).
The Armagh Observatory and Planetarium is funded by the Northern Ireland Department for Communities.
CAT thanks Churchill College for his Fellowship.




\bibliographystyle{mnras}
\bibliography{references}








\bsp	
\label{lastpage}
\end{document}